# The Impact of Organic Friction Modifiers on Engine Oil Tribofilms

Monica Ratoi[1]*, Vlad Bogdan Niste[1], Husam Alghawel[1], Yat Fan Suen[2] and Kenneth Nelson[2]

[1] Faculty of Engineering and the Environment, University of Southampton,
Highfield Campus, University Road, SO17 1BJ, Southampton, UK
[2] Chevron Oronite Company, Richmond, California, USA

*Corresponding author: M.Ratoi@soton.ac.uk

**Abstract**

Organic friction modifiers (OFMs) are important additives in the lubrication of machines and especially of car engines where performance improvements are constantly sought-after. Together with zinc dialkyldithiophosphates (ZDDPs) antiwear additives, OFMs have a predominant impact on the tribological behaviour of the lubricant. In the current study, the influence of OFMs on the generation, tribological properties and chemistry of ZDDP tribofilms has been investigated by combining tribological experiments (MTM) with in-situ film thickness measurements through optical interference imaging (SLIM), Alicona profilometry and X-ray photoelectron spectroscopy.

OFMs and antiwear additives have been found to competitively react/adsorb on the rubbing ferrous substrates in a tribological contact. The formation and removal (through wear) of tribofilms are dynamic processes which result from the simultaneous interaction of these two additives with the surface of the wear track. By carefully selecting the chemistry of OFMs, the formulator can achieve lubricants that generate ZDDP antiwear films of optimum thickness, morphology and friction according to the application-specific requirements.

## 1. Introduction

Modern engine oils contain a large number of additives, but the most influential on the tribological performance of the lubricant are antiwear zinc dialkyldithiophosphates (ZDDPs) and friction modifiers (FMs).

ZDDPs have been used in engine oils for more than 70 years and are probably the most successful antiwear additives ever discovered. In addition to being good antioxidants [1], they reduce wear by rapidly forming relatively thick, sacrificial boundary tribofilms with hardness characteristic to soft polymeric materials [2, 3].

The generation of ZDDP tribofilms on steel surfaces and their nature has been the main focus of published research. It has been shown that they form only on the rubbing tracks and are dependent on temperature [1]. On steel, tribofilms can rapidly grow to a thickness >100 nm and have an uneven, pad-like distribution (typically 5 to 20 μm across), separated by deep fissures. A two layer model was proposed, where a soft polyphosphate film covers a more rigid oxide-sulphide layer chemisorbed onto the steel surface [2]. The chemistry of the ZDDP oil solution was reported to affect the chemical make-up, thickness and mechanism of formation of the antiwear film [4]. The antiwear films formed by ZDDP have high boundary friction coefficients in the range of 0.11 to 0.14, which are maintained up to much higher sliding speeds than is normally the case. This is thought to be due to the unusual morphology of the reaction film [5]. To reduce the high boundary friction of the ZDDP tribofilms, it is especially important to use efficient friction modifiers when formulating lubricants.

Friction modifiers (FMs) can improve lubricity and thus energy efficiency by reducing the coefficient of friction (COF) in the boundary, and in some cases, also in the mixed lubrication regimes [6]. A direct application of this can be found in engine oils, which contain FMs for fuel economy purposes.

Presently, there are two main types of FMs: organic friction modifiers and molybdenum compounds. Organic friction modifiers (OFMs) are surfactant-like molecules with long chains and polar groups (e.g. alcohol, amide, carboxylic acid and ester groups), which adsorb or chemically react on the polar metal surfaces to form dense monolayers (2 nm thick) or thick reacted viscous layers [7, 8]. Organic molybdenum compounds, such as molybdenum dithiocarbamate (MoDTC), generate nanosized single sheets of $MoS_2$ dispersed in a carbon [9] or $FeS_2$ [10] matrix. The $MoS_2$ sheets facilitate sliding and thus lower friction between rubbing asperities.

Both antiwear and friction modifier additives work by reacting or adsorbing on the lubricated contact surface. Therefore, since these two types of additives are used together, it is important to understand the mechanism of action for generating the tribofilm and how these additives interact during this process. Unfortunately, at the present there is no systematic understanding of the interaction of FMs with other lubricant additives. For this reason, the selection of optimal additive combinations for lubricant formulation primarily depends upon trial and error or past experience, rather than knowledge of fundamental chemical interactions.

Previously published work investigating the interaction between ZDDP and FMs mainly focused on ZDDP and MoDTC. The reason for the high interest in MoDTC is that aside from being a FM, it is also one of the most efficient non-phosphorous antiwear additives [11], shown to synergize well with ZDDP to reduce wear and friction [10, 12-14]. However, the downside of using MoDTC alone or in combination with ZDDP in oils is the formation of abrasive $MoO_3$, which is conducive to high friction [15]. One study investigated the synergism between ZDDP and MoDTC and reported that there is a competitive adsorption on the rubbing steel surface between ZDDP and MoDTC, which results in a thinner ZDDP tribofilm [16].

The only published study which investigated OFMs besides MoDTC has proposed a different mechanism of action, in which FM additives form a friction-reducing film not on the ferrous surface, but on the zinc phosphate. Therefore, it was recommended for FMs to be tested and optimized for effectiveness on the ZDDP films [17].

The objective of the current study was therefore to explore the interaction between ZDDPs and OFMs and to determine whether this influence is susceptible to variations of the OFM additive type. To accomplish this objective, the work employed various techniques to study the growth kinetics, the physical (thickness, morphology) and tribological properties (friction and wear) and the chemical composition of tribofilms generated by ZDDP and three types of OFMs.

## 2. Experimental Section

The interaction between ZDDP and OFMs has been studied in three experimental steps. Firstly, ZDDP and three OFMs were separately added to a mineral base oil and their ability to form a tribofilm was investigated. Secondly, fully formulated oils containing all additives normally present in engine oils (ZDDPs, dispersants, detergents, antioxidants etc.) *except* OFMs were tested to study the influence of the other additives on ZDDP tribofilm formation. Finally, fully formulated engine oils (containing ZDDP, OFMs and other engine oil additives) were investigated.

The nine test fluids tested were supplied by Chevron Oronite Company and are shown in **Table 1**. The antiwear additive was a mixed primary/secondary alkyl ZDDP. Three types of OFMs were studied: a polyamidoalcohol (A), an alkenoic ester (B) and an oleylamine (C). The base oil (BO) is a group II mineral oil with a viscosity of 22.6 cSt at 40˚C and 4.6 cSt at 100˚C. Fluid 2 is a mixture of BO and ZDDP (0.5 wt. %) and fluids 3-5 are mixtures of BO and an OFM (A, B and C) at a 0.5 wt. % treat rate. The Baseline oil (BLO) is a fully formulated engine oil (5W20) containing BO, ZDDP and other additives but *without any* OFMs. Fluids 7-9 are mixtures of Baseline oil (BLO) and OFMs (A to C).

*Table 1: Lubricants tested*

1. BO
2. BO+ZDDP
3. BO+OFM A
4. BO+OFM B
5. BO+OFM C
6. BLO (BO+ZDDP + Other Additives but *NO* OFMs)
7. BLO+OFM A
8. BLO+OFM B
9. BLO+OFM C

Tribological tests were carried out in a Mini Traction Machine (MTM2) in a sliding-rolling ball-on-disc setup. This features a 3/4 inch ball and a 46 mm diameter disc, both made of AISI 52100 steel (hardness 750-770 HV). The root-mean-square roughness of both balls and discs is 11 ± 3 nm, resulting in a composite surface roughness of 16 nm. New specimens (balls and discs) were used for each test and were cleaned with solvents in an ultrasonic bath for 10 minutes prior to testing. The temperature was kept constant (100 °C) throughout the test. The applied load was 30 N, corresponding to an initial mean Hertz pressure of 0.94 GPa. The slide-roll ratio (SRR), defined as the ratio of the sliding speed $|u_b-u_d|$ to the entrainment speed $(u_b+u_d)/2$ (where $u_b$ and $u_d$ are the speed of the ball and the disc, with respect to the contact) was 150%. This slide roll ratio value was selected to be higher than in previously reported research [18] to accelerate the generation of the chemically reacted tribofilm, which is known to depend on the severity of the rubbing conditions [2]. The MTM2 is fitted with the 3D Spacer Layer Imaging Method (SLIM) attachment, which enables in situ capture of optical interference images of the tribofilms on the steel ball. From these images is calculated the tribofilm thickness generated during the test (Figure 1).

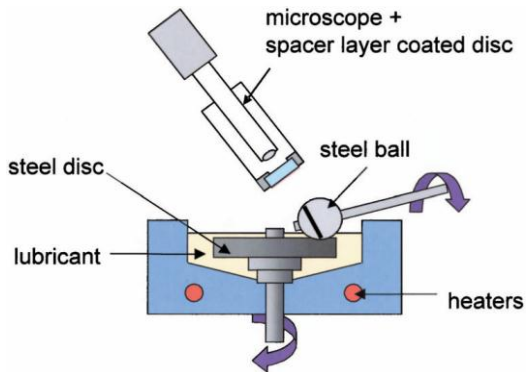

**Fig. 1** Diagram of MTM2-SLIM set up [19]

The tribological tests followed a routine which can be divided in three alternative stages repeated at fixed time intervals. First, the ball and disc were rubbed together at a fixed slow entrainment speed in mixed lubrication film conditions to generate a tribofilm on the ball and disc wear track ('*conditioning phase*'). This is followed by the '*Stribeck curve*' acquisition, in which friction was measured over a range of entrainment speeds at a fixed slide-roll ratio. The acquisition of data for the Stribeck curve started at the highest speed (1.5 m/s) and continued towards the lowest speed value (10 mm/s) to protect the formed tribofilm by avoiding its damage at low speeds in the boundary regime. Lastly, the motion was halted, the spacer layer-coated window was loaded against the ball track and an image was captured ('*tribofilm measurement*'). Table 2 summarizes the conditions used for the MTM2-SLIM tests in this study. Alicona Infinite Focus profilometry was used to study the thickness and morphology of the tribofilms across the wear track on the disc specimen after each test. The contact specimens (ball and disc) were made of the same material (AISI 52100) and have similar roughness values. Therefore, it is expected that both specimens form tribofilms of very similar composition and thickness values.

*Table 2: MTM – SLIM test conditions*

**Conditioning phase**

| | |
|---|---|
| Temperature | 40°C, 100°C |
| Load / Mean Hertz pressure | 30 N / 0.94 GPa |
| Entrainment speed | 0.1 m/s |
| Slide-roll ratio | 150 % |

**Stribeck curve phase**

| | |
|---|---|
| Temperature | 40°C, 100°C |
| Load / Mean Hertz pressure | 30 N / 0.94 GPa |
| Entrainment speed | 1.5 to 0.01 m/s |
| Slide-roll ratio | 150 % |

XPS measurements were performed using a Phi-Ulvac Quantera Scanning X-ray Microprobe. The excitation source for these characterizations was a monochromatized Al Ka (1476.6 eV) X-ray beam of 100 micron diameter. Points on and off the wear track were selected using secondary X-ray images of the surface including the wear track. The instrument spectral resolution was 1.1 eV. In depth profiling experiments, XPS characterizations were alternated with rastered $Ar^+$ beam milling. The milling rate was calibrated using a $SiO_2$/Si substrate with known oxide thickness, in accordance with standard techniques.

**3. Results and Discussion**

*ZDDP in Base Oil and BLO*

Figures 2 and 3 show a series of Stribeck friction curves measured during a 3-h rubbing test using BO+ZDDP and BLO fluids. They illustrate the dynamics of the ZDDP film growth throughout the mixed and boundary lubrication regimes. Figures 4 and 5 show series of optical interference images of the tribofilms formed on the steel ball by BO+ZDDP and BLO, respectively. The sliding direction in all these images is from left to right. The development of a patchy ZDDP antiwear chemical film on the wear track during rubbing is indicated by the dark areas. The interference images are used to calculate the average ZDDP tribofilm thickness in the central region of each image. These values are plotted against time in Figure 8.

The optical interference images and the Stribeck curves indicate that the morphology and thickness of the tribofilms generated by the two lubricants on the wear track, as well as their friction characteristics are significantly different. This indicates that the presence of other additives in the BLO lubricant influenced the ability of ZDDP to generate tribofilms.

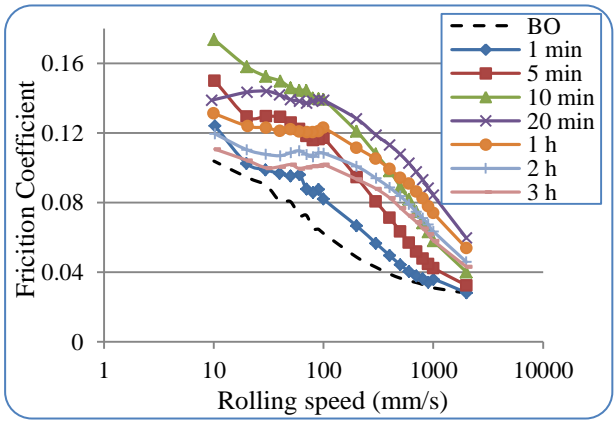

**Fig. 2** Series of Stribeck curves obtained during prolonged rubbing test with ZDDP+BO

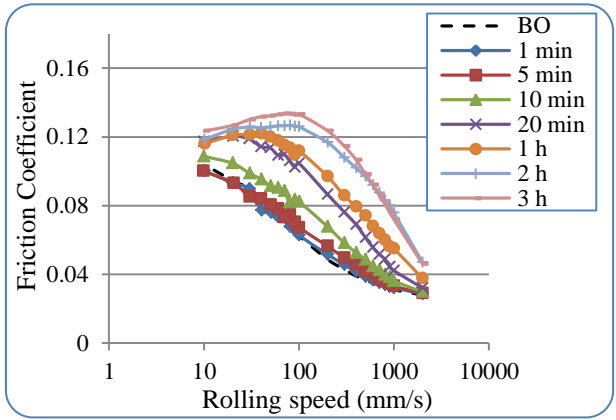

**Fig. 3** Series of Stribeck curves obtained during prolonged rubbing test with BLO

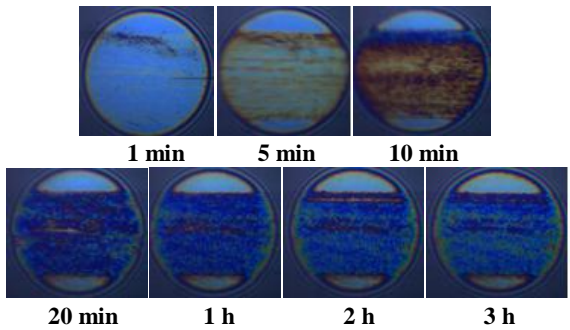

**Fig. 4** Optical interference images of the tribofilm formed by BO+ZDDP

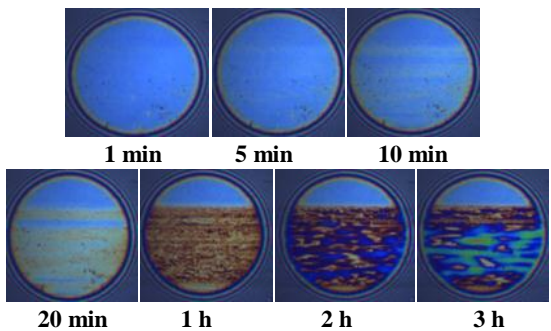

**Fig. 5** Optical interference images of the tribofilm formed by BLO
5

The formation of ZDDP films in rubbing contacts is a dynamic process, influenced by a number of testing conditions, such as temperature, time, load, slide-roll ratio etc. The value of the film thickness arises from a balance between the growth rate and the rate of removal determined by wear [20, 21].

Previously published research which employed the MTM tribometer showed that with 50% SRR, ZDDP tribofilms grow to a maximum of ~170 nm after 4 hours of rubbing [5, 17]. The optical interference images depict non-uniform films and the Stribeck curves show that friction continues to increase during testing in the mixed lubrication regime until it reaches a plateau.

The current study has employed similar testing conditions as the work reported above [5,17] except for the SRR which was significantly higher (150%). This SRR value resulted in the formation of thicker tribofilms (200+ nm) at a faster rate on one side, but also the removal through accelerated wear on the other. The tribofilms were smoother and more uniformly distributed, as shown by the SLIM images and the low friction values of the Stribeck curves measured after 1, 2 and 3 hours of rubbing.

This is in agreement with other published work, which reported that the morphology of ZDDP derived tribofilms evolves with rubbing time. The early formed pads were distinctly segregated and elongated along the rubbing lines, but after 1 h of rubbing, they started to break down into smaller pads forming a more uniform [2, 22], thinner and smoother film [22].

In the current study, the thickness of the evolved tribofilms, measured with SLIM (in-situ at preselected intervals during the 3-h test) and Alicona (at the end of the 3-h test), continued to grow after the first hour of testing but at a lower rate (Figure 8).

Figure 8 shows that the BO+ZDDP lubricant generated tribofilms faster (with a higher growth rate) than BLO. In the first 20 minutes the film shows an accelerated growth. After that, it continues to increase in thickness at a slower rate, up to approximately 200 nm and becomes more uniform.

To investigate the elemental composition and thickness of the tribofilms, XPS profiling using $Ar^+$ sputtering was employed on the wear tracks of the MTM disc specimens (for 3 h tribofilms). Figures 6 and 7 reveal the elemental composition against the tribofilm depth for BO+ZDDP and BLO fluids. The chemical composition of the tribofilm generated by BO+ZDDP is similar to other published results [22-25] and shows zinc polyphosphates and iron/zinc sulphides as the main constituents.

The thinner tribofilms formed by BLO, as compared to BO+ZDDP, imply that the other additives present in BLO interfere with the ability of ZDDP to generate antiwear chemical films. As Figure 5 shows, the tribofilms formed by BLO are composed of non-uniformly distributed patches, elongated along the sliding direction and this morphology is maintained throughout the 3-h testing. In the case of BO+ZDDP, the film becomes uniformly distributed during the first hour of rubbing and is maintained until the end of the test. As the BLO films developed in thickness, the friction increased in boundary and mixed lubrication regimes and after 3 hours of rubbing the BLO films displayed higher COF values than BO+ZDDP films.

The sputtering data shown in Figure 7 for the BLO film is quantitatively different from that of the BO+ZDDP film and it shows that calcium replaced zinc as the most abundant metal in the top layers of the tribofilm. The proportion of Zn in the tribofilm has decreased 6 times, while the amounts of phosphates and sulphides are overall 1-2% smaller than those found in the BO+ZDDP film. This behavior is consistent with the results reported by other published studies that investigated the effect of detergents and dispersants on ZDDP tribofilm formation [11, 18].

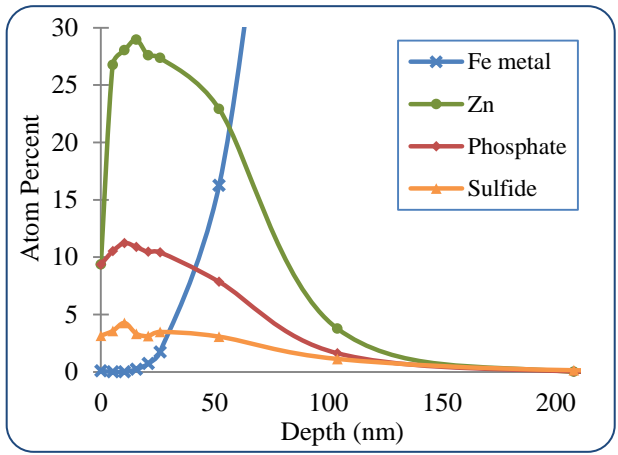

**Fig. 6** Composition of tribofilm generated during prolonged rubbing test with BO+ZDDP

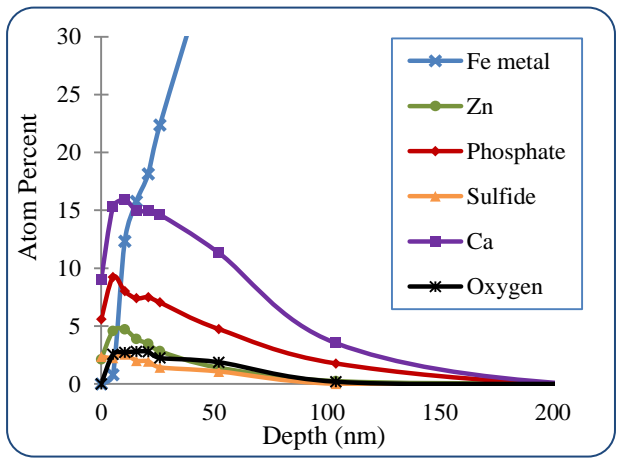

**Fig. 7** Composition of tribofilm generated during prolonged rubbing test with BLO

The sputtering depth data indicates that the thickness of the BO+ZDDP and BLO films are of approximately 150-200 nm, similar to the results measured with SLIM and Alicona.

*OFMs in Base Oil*

OFMs can reduce friction by physically adsorbing or, in some cases, chemically reacting with the steel substrate under rubbing conditions to form adsorbed or reacted films. OFMs that react with the steel substrate may have a chemical reactivity similar to the ZDDP antiwear additives and therefore, significantly interfere with the ZDDP film growth. In order to investigate the ability of studied OFMs to generate chemical tribofilms and their effectiveness, three-hour rubbing tests were carried out using OFM solutions in base oil.

Figure 8 shows the thickness and growth kinetics of tribofilms generated by OFMs in BO (compared to BLO and BO+ZDDP), measured with MTM2-SLIM and Alicona. BO+ZDDP forms thick films of approximately 200 nm at a fast rate, which follow an almost linear increase during the first hour of conditioning. OFMs B and C rapidly generated relatively thick (110 nm) reacted tribofilms. OFM A formed very thin reacted films that were difficult to quantify with SLIM or Alicona profilometry.

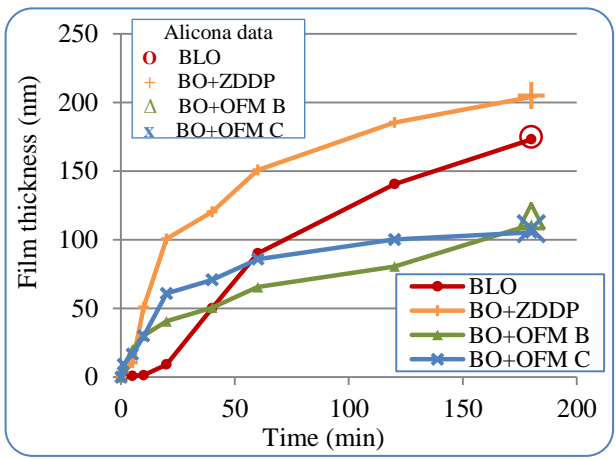

**Fig. 8** Film thickness with rubbing time for BO+OFMs / ZDDP using MTM2-SLIM and Alicona profilometry

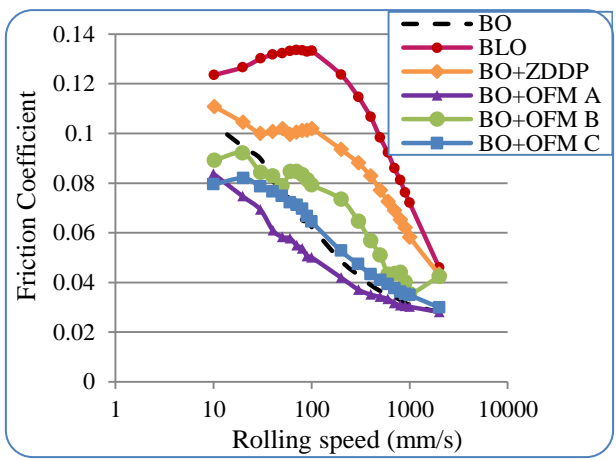

**Fig. 9** Stribeck curves for BO+OFMs / ZDDP after 3 h rubbing

Figure 9 shows the Stribeck curves for the OFMs measured after three hours of rubbing, compared to BO and BLO. BLO displayed the highest COF values in the boundary and mixed lubrication regimes. As discussed earlier, this can be explained by the rapid formation of a rough and irregular ZDDP tribofilm. All three OFMs were very efficient at reducing boundary friction as compared to BO, but their behaviour differed in the mixed lubrication regime. OFM A did not form a reacted tribofilm but was efficient at reducing friction throughout the whole speed range. In an opposite fashion, OFM B generated a reacted tribofilm similar to the ZDDP antiwear additive and was less effective at reducing friction in the mixed regime in comparison to BO.

*OFMs in Baseline Oil*

To investigate the extent to which OFMs interfere with the ZDDP film growth, each of the OFMs were separately added to BLO to make fully formulated oils. Film thickness measurements after 180 minutes of rubbing, shown in Figure 10, indicate that all OFMs suppressed the formation of ZDDP films by competing for the reaction sites. OFM B markedly reduced the formation of the ZDDP film, while OFM C and OFM A did so to a smaller extent. In the case of OFM C, the tribofilm started to form faster than in the absence of OFMs (BLO film).

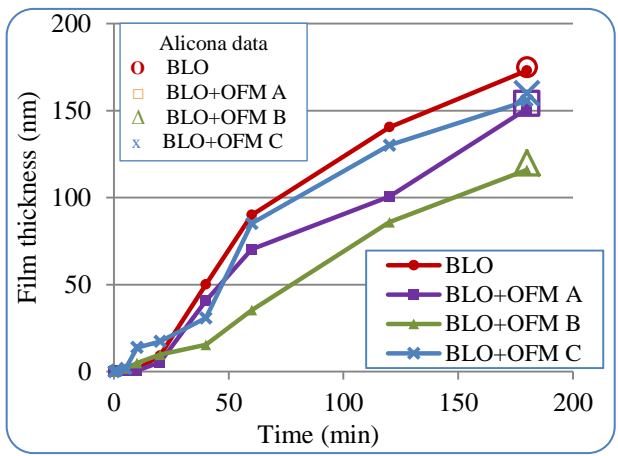

**Fig. 10** Film thickness with rubbing time for BLO+OFMs using MTM2-SLIM and Alicona profilometry

Stribeck curves after 180 minutes of rubbing are shown in Figure 11. All three OFMs reduced boundary friction. The most effective friction reduction in both, mixed and boundary regimes was obtained with BLO+OFM B, which generated the thinnest tribofilm. OFM C is effective at reducing only boundary friction, while the COF at high speeds (>100 mm/s) is comparable to BLO. The lubricants containing BLO+OFM C and BLO form films of comparable thickness, which may explain the similar COF at these speeds.

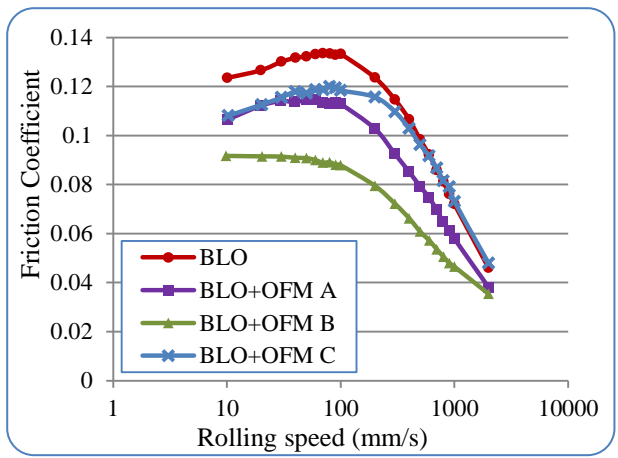

**Fig. 11** Stribeck curves for BLO+OFMs after 3 h rubbing

Out of the three OFMs, BLO+OFM B was the most efficient at reducing friction, but generated thinner tribofilms, while BLO+OFM C formed the thickest tribofilms, characterized by high COF values. BLO+OFM A generated relatively thick tribofilms at a slower rate and provided moderate friction reduction in the mixed regime. The friction and tribofilm thickness results for OFMs+BLO follow the same trend as the results generated by OFMs+BO.

The COF values for all the fluids tested at low, intermediate and high entrainment speeds (10 mm/s, 100 mm/s and 1000 mm/s) are shown in Table 3.

The antiwear properties of the formed tribofilms were assessed by measuring the wear track width on the disc specimens at the end of the three hours test, using Alicona profilometry. The 3 hours MTM test led to the formation of thick tribofilms above the surface level which made the calculation of wear loss not possible. The wear scar width results, presented in Table 3 are within a narrow range. From the three fully formulated fluids containing OFMs, the lowest wear is attained with BLO+OFM B and the highest with BLO+OFM C.

Both OFM B and OFM C generated own tribofilms in BO and when added to BLO, the tribofilm thickness was 90-120 nm for the former and 160-170 nm for the latter, while the amount of ZDDP species was 9% and 16%. As the ZDDP tribofilm composition and thickness were modified by the action of the two OFMs it is expected that the mechanical properties (i.e. hardness) should also differ. This explains why BLO+OFM C which forms thicker tribofilms also shows a larger wear scar than BLO+OFM B.

*Table 3: Comparison of tribofilm thickness, COF, wear scar and % ZDDP species (Zn, phosphate and sulfide) for all tested lubricants*

|  | Tribofilm thickness (nm) |  | COF |  |  | Wear scar (μm) | % ZDDP species at film surface (from XPS) |
| --- | --- | --- | --- | --- | --- | --- | --- |
|  | SLIM/Alicona | XPS | 10 mm/s | 100 mm/s | 1000 mm/s |  |  |
| BO | - | - | 0.10 | 0.06 | 0.05 | 255 | - |
| BO + ZDDP | 200 | 200 | 0.11 | 0.10 | 0.06 | 260 | 44 |
| BLO | 170 | 200 | 0.12 | 0.13 | 0.06 | 250 | 17 |
| BO + OFM A | - | - | 0.08 | 0.05 | 0.03 | - | - |
| BLO + OFM A | 150 | 90 | 0.11 | 0.11 | 0.04 | 250 | 5 |
| BO + OFM B | 110 | - | 0.09 | 0.08 | 0.04 | - | - |
| BLO + OFM B | 120 | 90 | 0.09 | 0.09 | 0.04 | 240 | 9 |
| BO + OFM C | 110 | - | 0.08 | 0.06 | 0.03 | - | - |
| BLO + OFM C | 160 | 170 | 0.11 | 0.12 | 0.05 | 255 | 16 |

Figures 12, 13 and 14 show the variation in the elemental composition against the depth of etching for BLO+OFM A, BLO+OFM B and BLO+OFM C.

As summarized in Table 3, the tribofilm thickness results supplied by the XPS sputtering data are relatively similar to the values measured with SLIM/Alicona except for BLO+OFM A where the XPS data indicate a lower value. The percentage of ZDDP species for the BLO+OFM A is the smallest, despite the fact that OFM A did not generate a measurable tribofilm in BO and was expected to interfere less with the ZDDP film formation. These results for BLO+OFM A may be attributed to the inconsistencies in the distribution of the tribofilm on the wear track.

Out of the three OFMs, OFM C generated the thickest tribofilm, similar to the BLO film in thickness and composition (ZDDP components in similar percentage). This indicates that OFM C does not interfere and it may even facilitate the ZDDP reaction with the substrate. OFM B, which along with OFM C generated tribofilms in BO, formed the thinnest tribofilm in BLO with a small percentage of ZDDP species. Based on these results, it can be concluded that OFM B hinders the reaction of ZDDP with the substrate.

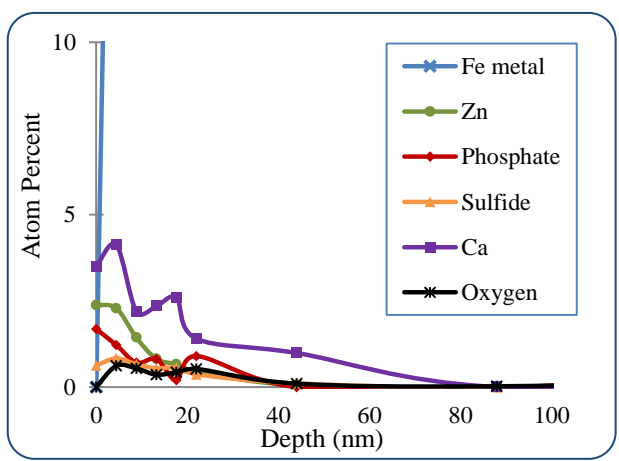

**Fig. 12** Composition of tribofilm generated during prolonged rubbing test with BLO+OFM A

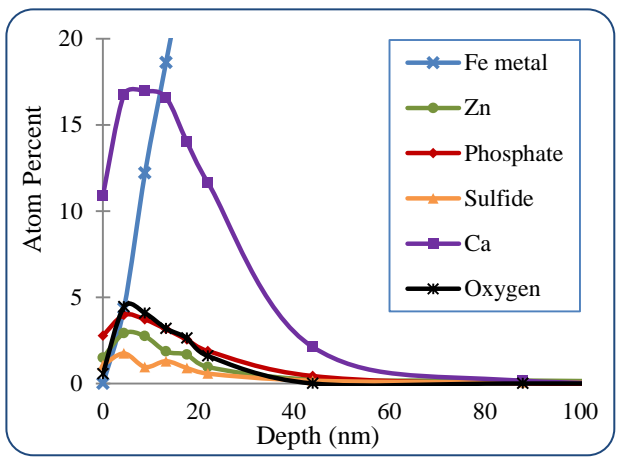

**Fig. 13** Composition of tribofilm generated during prolonged rubbing test with BLO+OFM B

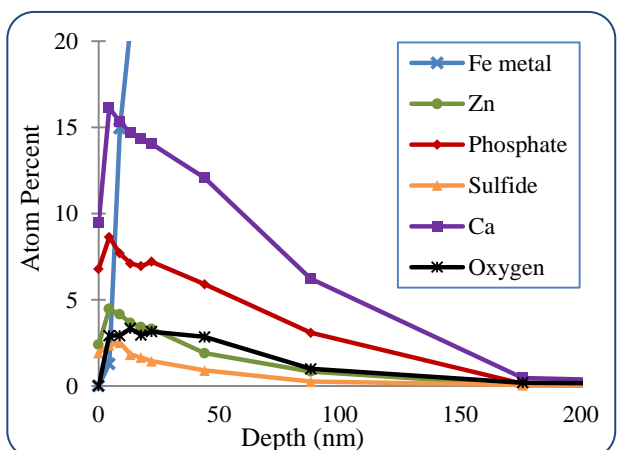

**Fig. 14** Composition of tribofilm generated during prolonged rubbing test with BLO+OFM C

The amount of ZDDP species (the summation of zinc, phosphates and sulfides) present in the tribofilms was found to decreases in the order: BO+ZDDP (44%) > BLO (17%) > BLO+OFM C (16%) > BLO+OFM B (9%) > BLO+OFM A (5%) and it follows the same trend as the XPS tribofilm thickness results.

## 4. Conclusions

This study has investigated the influence of three organic friction modifiers (OFMs) of different chemistries on the ZDDP tribofilm formation, composition and properties. The results show that the tribological properties of the fully formulated engine oils significantly depend on their composition.

Tests carried out with fully formulated oils without OFMs (BLO) showed that the generated ZDDP tribofilm is non-uniform and thinner (as shown by optical interference images), rich in calcium and poorer in zinc concentration (XPS spectra) than the BO+ZDDP films.

Whether in the presence or absence of other additives, OFMs can greatly influence the reaction of the ZDDP antiwear additive with the steel substrate and consequently, the kinetics, thickness, composition and tribological properties of the tribofilm generated in the rubbing contact.

Depending on their chemical composition, OFMs can react with the wear track in rubbing contacts to generate own tribofilms (as in the case of BO+OFM B and BO+OFM C) which have kinetics and thickness comparable to the ZDDP tribofilms (BO+ZDDP).

In a tribological contact, the formation and removal (through wear) of tribofilms are dynamic processes controlled by the interaction between the surface active additives and the lubricated contact surfaces. This is especially the case of antiwear and OFM additives, which compete to react/adsorb on the rubbing ferrous substrates. The generation of tribofilms is the result of the simultaneous interaction of these two additives with the surface of wear track.

When mixed into the fully additized oil (BLO), the three OFMs of various chemistries influenced the tribofilm generation and properties in particular ways. OFM B produced thin tribofilms (110 nm) and was very efficient at reducing friction (COF=0.09), while OFM A and C generated thick tribofilms (160 nm) but reduced friction to a lesser extent (COF=0.11).

Despite their different behaviour influenced by chemistry, all three OFMs have potential value for tribological applications. In automotive transmission, where emphasis is placed more on wear protection than friction reduction, OFM C and OFM A, which generate thicker tribofilms could be more useful. Engine oils require high levels of friction reduction and fuel efficiency and could be formulated with friction modifiers similar to OFM B.

The investigation of the effect of OFMs chemistry on ZDDP antiwear film performance will guide the formulator in selecting OFMs according the application-specific requirements.

**Acknowledgements**

The authors wish to thank Chevron Oronite Company for sponsoring this study and carrying out the XPS analysis.

**Conflict of interest**

The authors declare no competing financial interest.